\documentclass[
 reprint, 
superscriptaddress,
 amsmath,amssymb,
 aps, physrev,
]{revtex4-2}

\usepackage{graphicx}
\usepackage{dcolumn}
\usepackage{bm}

\begin{document}

\title{\textbf{Confinement controls bacterial spreading at all scales} 
}%

\author{Renaud Baillou}
\affiliation{%
Laboratoire PMMH-ESPCI Paris, PSL Research University, Sorbonne Universit\'e and Denis Diderot, 7, quai Saint-Bernard, Paris, France.
}%

\author{Marta Pedrosa Garc{\'i}a-Moreno}
\affiliation{%
CY Cergy Paris Universit{\'e}, Laboratoire de Physique Th{\'e}orique et Mod{\'e}lisation, UMR 8089, 95302 Cergy-Pontoise, France.
}%

\author{Quentin Guigue}
\affiliation{%
Laboratoire PMMH-ESPCI Paris, PSL Research University, Sorbonne Universit\'e and Denis Diderot, 7, quai Saint-Bernard, Paris, France.
}%

\author{Solene Meinier}
\affiliation{%
Laboratoire PMMH-ESPCI Paris, PSL Research University, Sorbonne Universit\'e and Denis Diderot, 7, quai Saint-Bernard, Paris, France.
}%

\author{Gaspard Junot}
\affiliation{%
Laboratoire PMMH-ESPCI Paris, PSL Research University, Sorbonne Universit\'e and Denis Diderot, 7, quai Saint-Bernard, Paris, France.
}%

\author{Thierry Darnige}
\affiliation{%
Laboratoire PMMH-ESPCI Paris, PSL Research University, Sorbonne Universit\'e and Denis Diderot, 7, quai Saint-Bernard, Paris, France.
}%

\author{Fernando Peruani} \email{Contact author: fernando.peruani@cyu.fr}
\affiliation{%
CY Cergy Paris Universit{\'e}, Laboratoire de Physique Th{\'e}orique et Mod{\'e}lisation, UMR 8089, 95302 Cergy-Pontoise, France.
}%

\author{Eric Cl{\'e}ment} \email{Contact author: eric.clement@upmc.fr}
\affiliation{%
Laboratoire PMMH-ESPCI Paris, PSL Research University, Sorbonne Universit\'e and Denis Diderot, 7, quai Saint-Bernard, Paris, France.
}%
\affiliation{%
Institut Universitaire de France (IUF).
}%

\date{\today}

\begin{abstract}

Navigation of microorganisms is controlled by internal processes ultimately sensitive to mechanical or chemical signaling encountered along the path. In many natural environments, such as porous soils or physiological ducts, motile species alternate between bulk and surface  motion displaying in each case, distinct kinematics. This inherent complexity is key to many practical biological and ecological issues involving  spreading and contamination, essential for understanding the spatiotemporal structuring of populations in their environment. However grasping the interplay between geometrical confinement and kinematics driven by internal biological responses remains poorly understood from a physical and biological standpoint.
Here, we address this question through experimental and theoretical analysis in the heuristic situation of two parallel confining surfaces. We track wild-type {\it E. coli} — a model peritrichous flagellated bacterium — in 3D over extended periods of time. We obtain the first experimental measurements of the emerging diffusivity and bulk/surface residence times as a function of confinement height and the specific chiral kinematics at surfaces. All experimental results are quantitatively reproduced, without parametric adjustment, by a non-Markovian stochastic (BV) model that incorporates the internal biochemical memory carried by a phosphorylated protein switching the motor rotation.
By matching the results with a Markovian (memoryless) companion model, we derive an analytical expression for the diffusivity and demonstrate how confining walls influence microbial long-range dispersion. This approach also provides a general conceptual basis for understanding how microorganisms navigate complex environments, in which their movement alternates between bulk and surfaces.

\end{abstract}

\maketitle


In free space, the three-dimensional (3D) exploration of wild-type \textit{E. coli} follows a run-and-tumble (R\&T) process, characterized by straight runs interrupted by abrupt changes in the moving direction, called tumbling events~\cite{Berg2004}. During the run phase, all motors driving the helical flagella rotate counterclockwise (CCW), and due to hydrodynamic interactions a bundle of flagella is formed. The assembly of CCW rotating flagella propels the bacterium forwards for a specific time, referred to as a \textit{run time}. A tumble occurs when one or more motors switch to a clockwise (CW) rotation, leading to the bundle disassembly and a reorientation of the bacterium body. Once all motors resume a CCW rotation, the bundle reassembles, and the bacterium begins a new run phase~\cite{Darnton2007}.  

Pioneering studies by Brown and Berg in the 1970s~\cite{Berg1972, brown1974temporal} suggested that, in the absence of chemical gradients, the distributions of both run and tumble times exhibit an exponential decay. Consequently, it was assumed that memoryless, stochastic, Poisson processes govern the time series of run and tumble events~\cite{Berg1972}. Under this assumption, the diffusivity that characterizes the emerging 3D random walk, scales as $ V^2 \tau_r$, where $V$ is the swimming speed, and $\tau_r$ is the mean run time. Since then, this Poissonian R\&T framework has formed the basis for numerous theoretical and numerical studies on microbial transport properties and collective self-organization processes~\cite{Elgeti_2015}.  

However, later experiments have challenged this initial analysis that assumes an exponential run-time distribution. Measurements of CCW rotation durations in tethered \textit{E. coli} showed non-exponential, long-tailed distributions~\cite{Korobkova2004}, putting forwards the existence of a behavioral memory~\cite{Tu2005,Emonet2008}. Figueroa-Morales et al.~\cite{figueroa20203d} demonstrated that the temporal sequence of swimming persistence times can be described by a model involving fluctuations of an internal variable, the phosphorylated CheYP protein, leading to behavioral memory and large log-normal run time distributions. Similar log-normal distributions were observed for bacterial residence times on flat surfaces~\cite{Junot2022}, contrasting further with the original assumptions by Berg et al.~\cite{Berg1972}. Importantly, the presence of very long runs can significantly affect bacteria dispersion in confined flows, as demonstrated by the facilitation of backflow contamination in narrow channels~\cite{figueroa2020coli}.  However, the question of the emergence of large-scale motile bacterial dispersion in a quiescent situation under confinement is still largely unraveled.

In contrast to the R\&T motility pattern observed in the bulk, near planar solid surfaces, bacteria exhibit a fundamentally different type of motion. Hydrodynamic interactions force bacteria to swim in circular CW trajectories~\cite{Frymier1995,Lauga2006,Berke2008,Lemelle2013}. Consequently, the surface diffusivity becomes significantly different from the bulk diffusivity, explicitly depending on the average radius of curvature of such circular trajectories~\cite{otte2021statistics} as well as on bacterial surface adhesion properties~\cite{perez2019bacteria}, while surface residence times are controlled by run time statistics~\cite{Junot2022}.  

Natural environments such as soil crevices \cite{Ranjard_2001}, or physiological ducts present complex landscapes where planktonic bacteria populations\cite{Guasto_2012} exploring their surroundings cannot swim purely in 3D or purely in 2D. Instead, their motion involves frequent transitions between 3D swimming in the bulk and quasi-2D motion near the surfaces. These transitions, as we explain here, play a crucial role in shaping large-scale bacterial transport properties, and yet, despite their relevance, their impact on the emerging diffusivity has remained, to date, largely unraveled~\cite{Biondi_1998,Lynch_2022,Raza_2023}. 
Here, we investigate bacterial transport in a simplified geometry consisting of two parallel surfaces separated by a distance $H$. We focus on the emergent bacterial diffusivity $D$ and show that $D$ explicitly depends on the confinement height $H$, being a function of the ratio between the average time spent in the bulk and on surfaces. 
We explain how the surface kinematics and an internal behavioral memory contribute to the emerging transport properties. Furthermore, we argue the observed scaling between $D$ and confinment height $H$ is universal for microswimmers navigating in confined spaces. These findings represent a critical step toward a comprehensive understanding of microorganisms exploration in natural environments.

\section{Bacterial spreading under confinement}

\begin{figure}[t!]
\centering
\includegraphics[width=\linewidth]{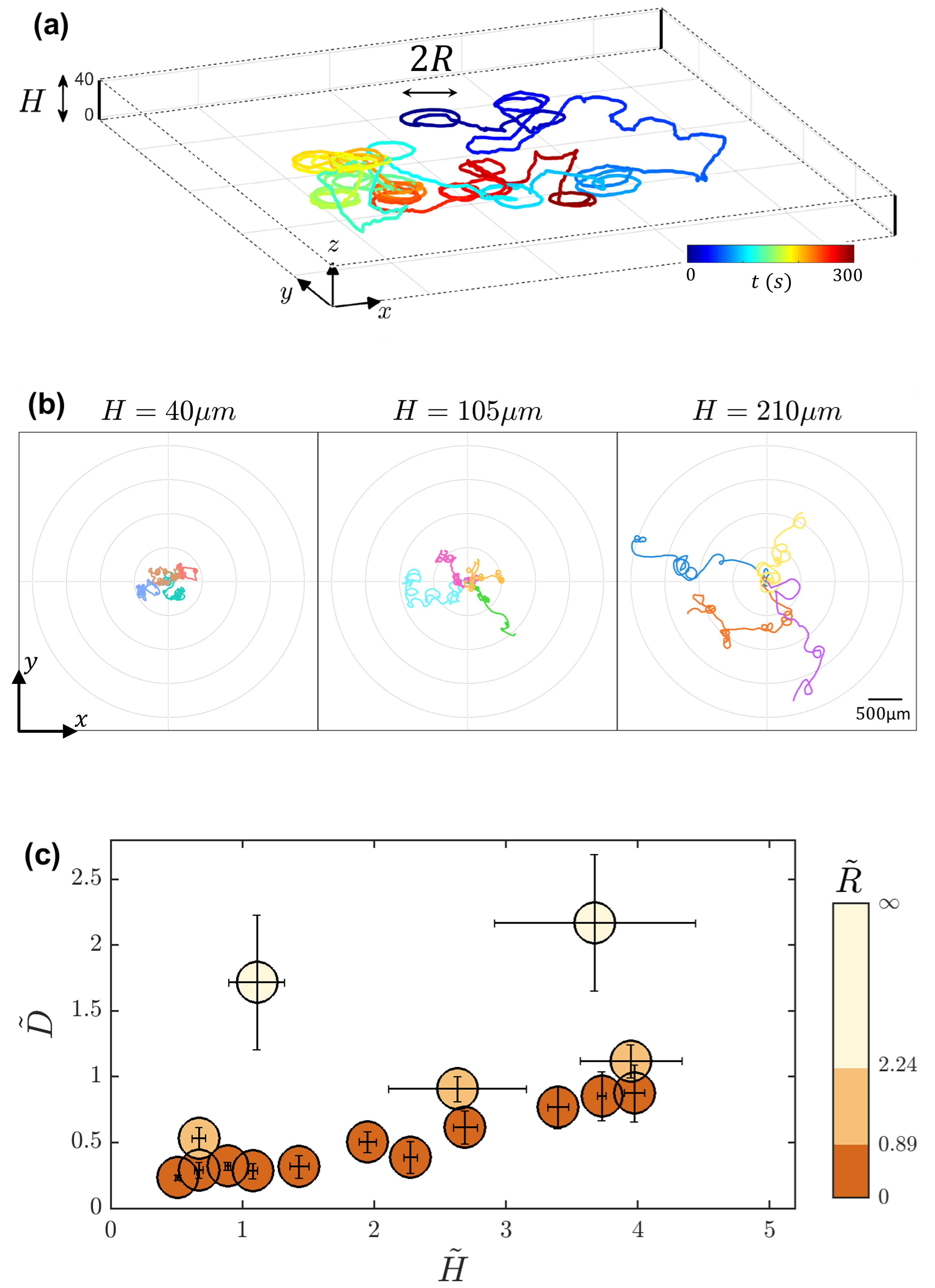}
\caption{  
\textbf{Spreading of motile wild-type \textit{E. coli} bacteria under confinement.}  
\textbf{(a)} 3D trajectory of a bacterium confined between two parallel glass plates ($H = 40\,\mu\text{m}$) tracked for $T = 300\,\text{s}$. Time is color-coded along the trajectory. The bacterium exhibits circular motion near the surface with a radius of gyration $R$. 
\textbf{(b)} Top view of typical bacterial trajectories, in different colors, over $T = 200\,\text{s}$ for varying confinement heights ($H = 40\,\mu\text{m}$, $105\,\mu\text{m}$, $210\,\mu\text{m}$), illustrating the effect of confinement on the spreading. Each trajectory is characterized by a swimming speed $V$, a surface radius of gyration $R$, a and lateral diffusion coefficient $D$.  
\textbf{(c)} Dimensionless diffusivity $\tilde{D} = D / (0.5V^2 \tau_r)$ as a function of dimensionless confinement height $\tilde{H} = H / (V \tau_r)$. 
The 92 tracks are categorized based on their dimensionless radii of gyration $\tilde{R} = R / (V \tau_r)$, in a range corresponding to the color bar.
Each data point represents an average over 6 tracks, with error bars indicating the standard error. 
The  diffusivity $\tilde{D}$, that quantifies the lateral spreading, increases both  with the confinement height $\tilde{H}$ and the radius of gyration $\tilde{R}$.  
}
\label{fig1}
\end{figure}
\begin{figure*}[t!]
\centering
\includegraphics[width=\linewidth]{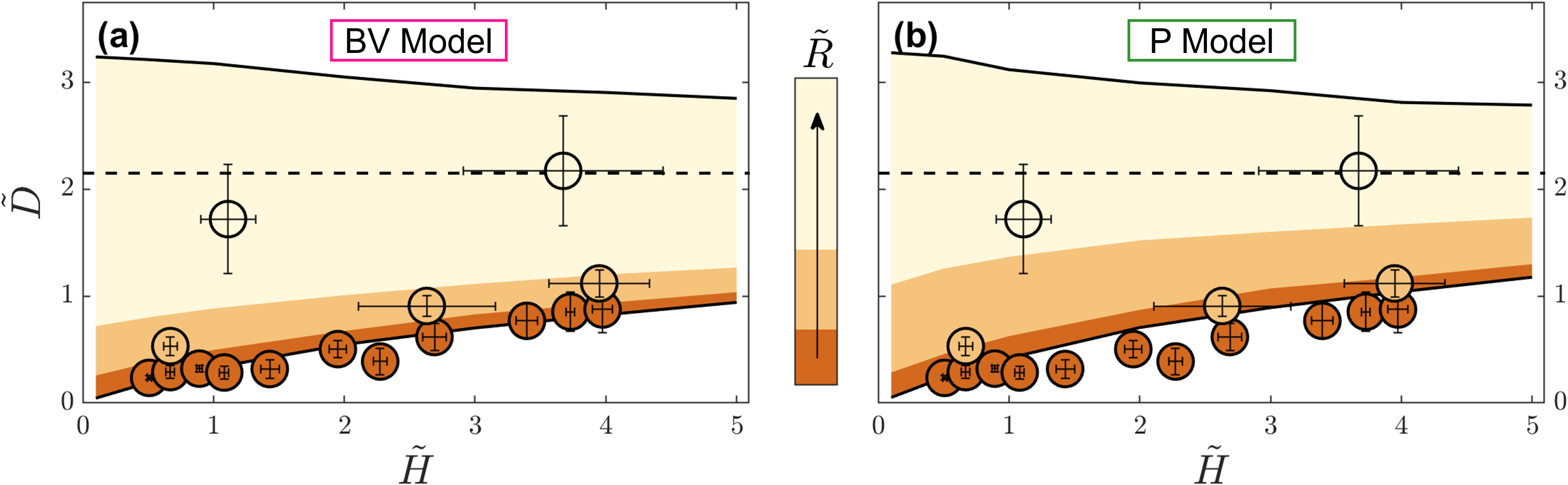}
\caption{ \textbf{ 
Spreading diffusivity $\tilde{D}$ vs confinement height $\tilde{H}$: direct comparison between experiments and numerical models.} Panel (a) corresponds to the BV Model, and panel (b) to the P Model, which follows a Poisson process of mean run times $\tau_r^P = 6.74\,\text{s}$, chosen to match the BV Model  boundless asymptotic limit $\Tilde{D_{\infty}}=2.148$ (dashed line). Experimental data are represented by colored symbols, while model predictions are shown as colored regions corresponding to the ranges of radii of gyration of Fig.~\ref{fig1}c. Bottom and top solid lines in panel (a) and (b) correspond to model predictions for resp. $\tilde{R}=0$ and $\tilde{R}\to\infty$. Note that the non-Markovian BV Model agrees quantitatively with the experimental data without requiring any fitting parameter. Its memoryless companion, the adjusted P Model, provides a simpler semi-quantitative description.}
\label{fig2}
\end{figure*}

Swimming bacteria -- wild-type motile {\it E. coli} [Appendix A] -- are placed in a minimal motility medium between two glass slides in a large pool of diameter around $1\,\text{cm}$ and separated by a confinement height $H$ varied between $40\,\mu\text{m}$ and $220\,\mu\text{m}$. 
The swimming microorganisms are observed via an inverted epifluorescence microscope and tracked in three dimensions using an in-house Lagrangian tracking device~\cite{Darnige2017lagrangian}. This tracking method was successfully implemented to monitor {\it E. coli} swimming in quiescent fluids~\cite{figueroa20203d}, Poiseuille flows, both in the bulk and at the surfaces~\cite{junot2019swimming,Junot2022}. A total of 92 tracks were collected, each lasting between $200\,\text{s}$ and $300\,\text{s}$, providing around 30 tracks for each confinement height $H=\{50\pm15\,\mu\text{m; }100\pm5\,\mu\text{m; }210\pm10\,\mu\text{m}\}$. 
The time durations of these 3D tracks are enough to let the bacteria visit both the upper and the lower glass slides several times [Fig.~\ref{fig1}a],  even  for  the largest $H$ values [Fig.~S1,S2]. The projection of each trajectory on the X-Y plane displays straight trajectory sections in the bulk followed by circular parts related to the near-surface swimming kinematics [Fig.~\ref{fig1}b].
Each bacterium, labeled $\kappa$, is characterized by its kinematic parameters [Appendix B]: the mean swimming velocity $V^{\kappa}$, the mean surface radius of gyration $R^{\kappa}$ and the \( xOy \) lateral diffusion coefficient $D^{\kappa}$.
\subsection{Lateral diffusion}
For a 2D  lateral diffusion stemming from a 3D boundless R\&T exploration at a swimming velocity $V$, the emerging large-scale diffusion coefficient is given by $D_0 = V^2 \tau_r/2$, assuming a Poisson process of mean run time $\tau_r$ and instantaneous, fully random tumbles~\cite{Lovely1975}. The elementary plot representing for all trajectories, $D^{\kappa}$ as a function of $V^{\kappa}$ does not show any clear tendency supporting this simplistic view [Fig.~S3]. This paradoxical result indicates that the emerging diffusivity could indeed be strongly influenced by confinement, detailed surface kinematics, or even bacterial phenotypic diversity. Therefore, to account for the observed variability, we consider dimensionless quantities for each bacterium based on its persistence length $l^{\kappa}_p = V^{\kappa} \tau_r$, where $\tau_r=2.23$s is the global mean run time provided in the previously calibrated numerical model~\cite{figueroa2020coli, Junot2022}. It yields, for each track, a dimensionless confinement height $\tilde{H}^{\kappa}= H /l^{\kappa}_p$ and a dimensionless radius of gyration  $\tilde{R}^{\kappa}= R /l^{\kappa}_p$. For lateral diffusion, normalization by the diffusion constant $D_0^{\kappa}= (V^{\kappa})^2 \tau_r/2$ is considered for each track. 
Bacteria are then classified into three groups according to their rescaled radii of gyration : (i) $0 < \tilde{R}^{\kappa} < 0.89$, (ii) $0.89 < \tilde{R}^{\kappa} < 2.24$, and (iii) $\tilde{R}^{\kappa} > 2.24$.
In Fig.~\ref{fig1}c, for each $\Tilde{R}$ group, the dimensionless quantities $\tilde{D}$ for the diffusivity and $\tilde{H}$ for the confinement height are averaged over six successive tracks ordered by $\tilde{H}$. 
For a given radius of gyration, the lateral diffusion coefficient increases (almost linearly) with the confinement height. For a given confinement, its value is larger for larger radii of gyration.
We then demonstrate the combined influence of confinement and surface kinematics on the lateral dispersion process.
\subsection{Residence times}
Moreover, the Lagrangian 3D tracking technique allows for the measurement of the durations a bacterium spends continuously in the bulk or near the surfaces, referred to as \textit{residence times}, and intrinsically linked to the diffusivity.
To account for inherent experimental uncertainties, we employed a \textit{practical} definition of the surface and bulk residence times, then denoted $T^*_s$ and $T^*_b$, based on a robust double-threshold algorithm~\cite{Junot2022} [Appendix C]. The notation with an asterisk (*) contrasts with the \textit{ideal} residence times, denoted $T_s$ and $T_b$, where the surfaces can be simply defined as \( z = 0 \) or \( z = H \).
They will be discussed after the introduction of the models. 

\section{Simulating mixed 3D and 2D exploration kinematics}
\subsection{BV Model} We now show that all experimental observations can be quantitatively reproduced within the framework of a non-Markovian R\&T model called the behavioral variability (BV) model [Appendix D1] introduced by Figueroa-Morales et al.~\cite{figueroa20203d} and extended by Junot et al.~\cite{Junot2022} to handle surface landing and take-off. Here, we add to the model the characteristic circular kinematics observed at surfaces, parametrized by a radius of gyration $R$. 
In this model, the source of non-Markovian stochasticity comes from a slow internal variable $X(t)$ coined as the \textit{mood} \cite{figueroa20203d}.  It represents the concentration of CheY-P proteins near the motor, which are responsible for the motor rotation switches that trigger tumbling\cite{Wadhams2004}. Mood fluctuations first introduced by Tu et al.\cite{Tu2005}, are modeled using an Ornstein-Uhlenbeck process  of characteristic memory time $T_M=19$s, typically ten times shorter than the duration of our tracks. Then, at steady-state, the distribution of $X(t)$ is Gaussian. Thus, with the switching rate depending exponentially on $X(t)$, the run time distribution naturally comes out as log-normal.
Direct comparisons between this non-Markovian R\&T model calibrated earlier~\cite{figueroa2020coli} and the experiments are performed  without introducing any extra free fitting parameter [Appendix D2]. Diffusivity values obtained from the simulations confirm that confinement heights $\tilde{H}$ and radii of gyration $\tilde{R}$ indeed control the diffusion coefficient [Fig.~\ref{fig2}a]. All data are comprised between two limits: (i) the "sticking" model with $\tilde{R}=0$ and (ii) the "SRun" model with $\tilde{R}=\infty$. 
Notably, Villa-Torrealba et al.~\cite{Villa_Torrealba_2020} have studied the unbounded "SRun" limit in 2D, showing a strong contribution of long run-times to the slow convergence towards the diffusive limit.
Remarkably, the BV Model outcomes agree quantitatively -- without adding any free parameter --  with all the experimental measurements.
\begin{figure}[t!]
\centering
\includegraphics[width=\linewidth]{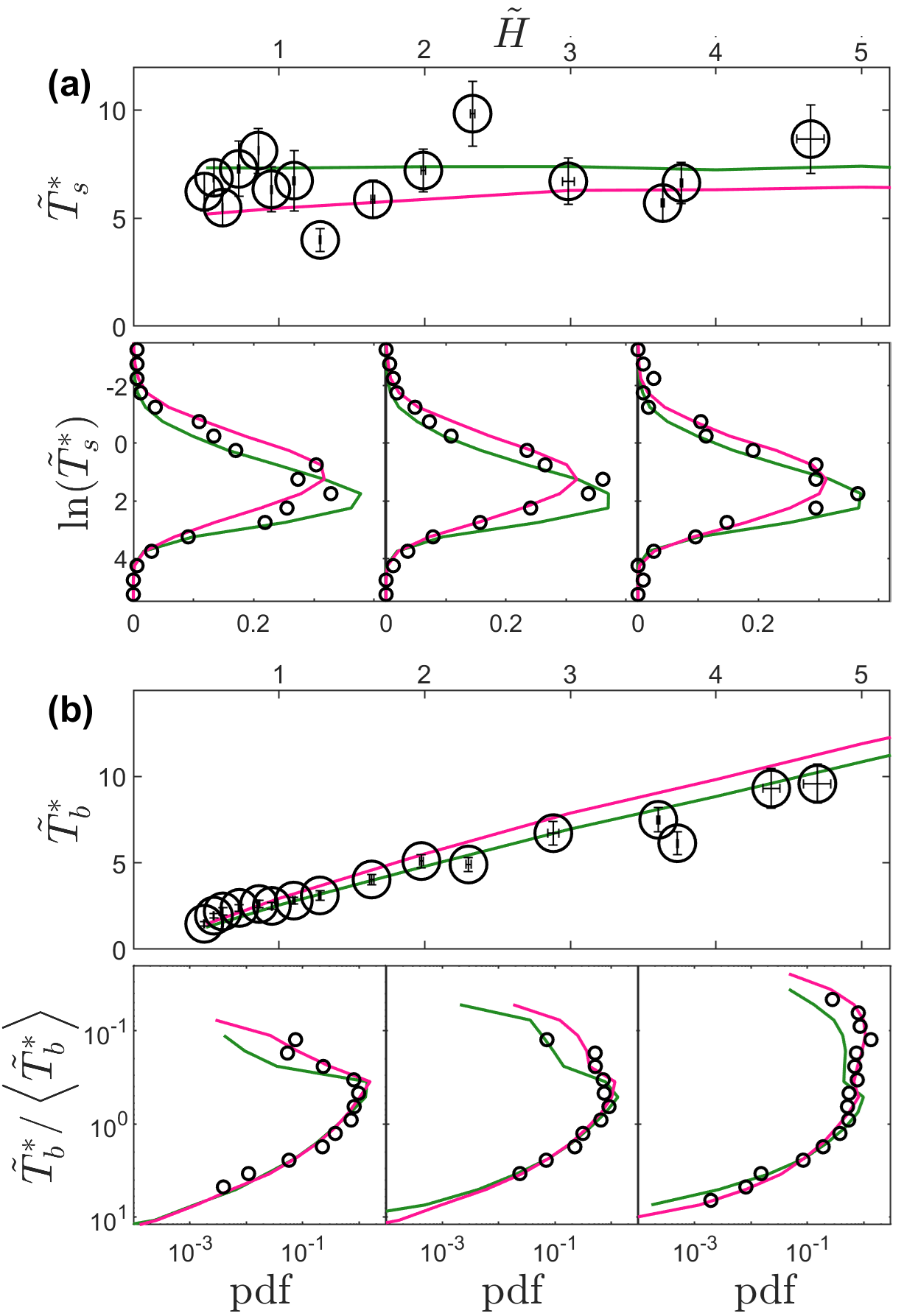}
\caption{ \textbf{ Experimental vs. numerical residence times: averages and distributions.} Experimental data are represented by black circles. Model predictions are pink lines (BV Model) and green lines (P Model). Error bars are the standard errors. The probability density functions (pdf) are shown for different confinement intervals: $0.5<\tilde{H}<1$, $1<\tilde{H}<2.5$, and $2.5<\tilde{H}<5$ (from left to right). 
\textbf{(a)} (top) Mean residence times at the surface $\tilde{T}^*_s={T}^*_s/\tau_r$ as a function of confinement height $\tilde{H}$, and (bottom) distribution of $\ln(\tilde{T}^*_s)$ for each confinement interval. 
\textbf{(b)} (top) Mean residence times in the bulk $\tilde{T}^*_b={T}^*_b/\tau_r$ as a function of confinement height $\tilde{H}$, and (bottom) distribution of the ratio $\tilde{T}^*_b/\left< \tilde{T}^*_b \right>$ for each interval, where $\left< \tilde{T}^*_b \right>$ represents the mean value of $\tilde{T}^*_b$ for each confinement range. 
For both models, residence times are in quantitative agreement with all the experimental data.}
\label{fig3}
\end{figure}
\subsection{P Model}
This good agreement with the BV-model prompts questions on the fundamental importance of the non-Markovian character entailed in the exploration process; or whether all the macroscopic outcomes would essentially be controlled by the presence of effectively large run time values. To clarify this important question, for example, in view of inherent difficulties to perform analytical calculations for non-Markovian models, we propose  to investigate the outcome of a companion Poissonian model: the P Model. This model provides the same asymptotic value for the 3D boundless diffusivity (see dashed lines in Fig.~\ref{fig2}), but with a run time distribution stemming from a memoryless Poisson stochastic process -- all other parameters being identical.
\subsubsection{Diffusivity in an unbounded 3D space}
Following the pioneering work of Brown and Berg ~\cite{brown1974temporal}, Lovely and Dahlquist~\cite{Lovely1975}  derived a general expression for the
diffusivity of a R\&T process taking place in an infinite, $d$-dimensional space. The calculation is based on the compound of free runs of length ${\mathbf{l}}_i$, lasting a time $\tau_{r,i}$ followed by a tumble resulting in an angular change $\theta_i$ lasting $\tau_{t,i}$ and taking place at rest. For runs performed at a fixed velocity $V$, considering a mean run time $\tau_r=\langle \tau_{r,i} \rangle_{i}$ and a mean tumble time $\tau_t=\langle \tau_{t,i} \rangle_{i}$, the diffusion coefficient is~\cite{Lovely1975} [Appendix D3]:
\begin{equation}
 D_{\infty} =  \frac{V^2 \tau_r}{d(1-\alpha)} ~f ~S ~[1+\alpha (1/S-1)] 
\label{LD}
\end{equation}
with the mean angular redirection coefficient $\alpha= \langle\cos{\theta_i}\rangle_i$, the fraction of time spent in the run phases, $f = {\tau_r}/({\tau_t+ \tau_r}) $,  and $S = {<\tau_{r,i}^2>_i}/({2\tau_r^2})$ a normalized second moment. 
Equation~\ref{LD} provides an excellent estimate of the lateral diffusivity in the $H\to\infty$ limit, even for the non-Markovian BV Model. 
Also, the  \( xOy \)-projected diffusivity for an isotropic R\&T process that occurs in an infinite 3D space is also $D_{\infty}$ with $d=3$.  
In dimensionless form, we obtain $\tilde{D}_{\infty} =D_{\infty}/D_0=2.148(5)$, quite far from most of the experimental outcomes.  
To ensure that the lateral diffusivity of the P Model has the same asymptotic value than the diffusivity obtained for the BV Model, we set -- considering that for a Poissonian process $S=1$ -- the  mean run time to $\tau_r^P = 6.74$s [Appendix D4]. 
Noticeably, this value is significantly higher than Berg's standard value~\cite{Berg2004} i.e. $\tau_r \approx 1$s, often considered in many theoretical or numerical quantitative  studies of the exploration and transport properties of motile {\it E. coli}. 
\subsubsection{Diffusivity on a 2D surface}
The diffusivity of bacteria constrained to swim near solid surfaces cannot be reduced to Eq.~\ref{LD} with $d=2$. 
The reason for this is that near-surface hydrodynamic interactions, as mentioned above, compel bacteria to swim in circles~\cite{Lauga2006}. 
It turns out that near-surface bacterial motion can be described with a simple chiral random walker model with characteristic radius of gyration $R$ and mean run time $\tau^P_r$~\cite{otte2021statistics}.   
The diffusivity $D_s(R)$ of such chiral walker model can be well approximated assuming Poissonian run times:
\begin{equation}
D_s(R) = \frac{ D_s(R\to\infty)}{1+(\frac{V\tau^P_r}{R})^{2} }
\label{D_sP}
\end{equation} 
This formula is strictly exact for all $R$ only for fully random reorientations.
Here we approximate $D_s(R\to\infty)= 3\tilde{D}_{\infty}/2= 3.22(2)$, very close from numerical results [Appendix D5]. 
Thus when $R\to\infty$, equation~\ref{D_sP} becomes equation~\ref{LD} with $d=2$ and when $R\to0$, $D_s$ vanishes.
\subsubsection{Interplay between 2D and 3D}
In short, for a memoryless Poisson model, there are two limiting situations where the diffusivity of P Model is known. When the bacterium moves in an unbounded 3D space ($H\to\infty$), the diffusivity is given by Eq.~\ref{LD}, and when the bacterium is constrained to move on a surface ($H\to0$), the diffusion coefficient is given by Eq.~\ref{D_sP}, which explicitly depends on the values of $R$. 
However, for any finite $H$, bacterial swimming constantly alternates between 3D and 2D motion, implying that asymptotically, the actual diffusion coefficient of this complex process is bounded by Eqs.~\ref{LD} and~\ref{D_sP}. 
We note that Pietrangeli et al.~\cite{Pietrangeli2024} have recently analyzed theoretically the impact of confinement on the diffusion coefficient of a Poissonian R\&T process, but  with four discrete directions and in a 2D slit geometry, with an interplay between 2D and 1D motion. Results obtained by Pietrangeli et al. also indicate that the  diffusivity -- which also remains bounded between two limiting cases -- is modulated by the confinement height.

Below, after comparing the numerical models with experiments and assessing the role of memory, we discuss an analytical model that interpolates between these limits and demonstrate that the lateral diffusivity is a function of $H$.  

\section{Models vs. experiments}
For the spreading diffusivity, the quantitative agreement between experiments and the P Model [Fig.~\ref{fig2}b] is of a lesser quality than with the BV Model [Fig.~\ref{fig2}a]. However, all qualitative tendencies expressing the quasi-linear dependence with the confinement height and the influence of surface kinematics are recovered and in a correct range of variations. 
Averages and distributions of residence times as a function of confinement are displayed in Fig.~\ref{fig3}. Experimental averages are computed over 65 successive residence times ordered by $\tilde{H}$, pooling all the radii together.
 For both models, bulk and surface residence times display a satisfactory quantitative agreement with the experimental outcomes. 
 The mean surface residence time is quite independent of the confinement [Fig.~\ref{fig3}a], whereas the mean bulk residence time increases almost linearly with the confinement height [Fig.~\ref{fig3}b].
Consistently with the run time distribution previously obtained by Junot et al.~\cite{Junot2022}, the distribution of surface residence times is quite large and can be described as a log-normal distribution [Fig.~\ref{fig3}a]. On the other hand, the distribution of bulk residence times [Fig.~\ref{fig3}b] decays exponentially and its shape varies, from one to two peaks, when the confinement height increases. This second peak essentially arises from bacteria coming back to their initial launching surface.
 Therefore, we have shown that the BV Model indeed displays a quantitative representation of the experimental situation. However, its Markovian companion model -- the P Model -- is sufficient to qualitatively capture the main features of the spreading process, provided a sufficiently large mean run time is used.
 %
\begin{figure}[t!]
\centering
\includegraphics[width=1\linewidth]{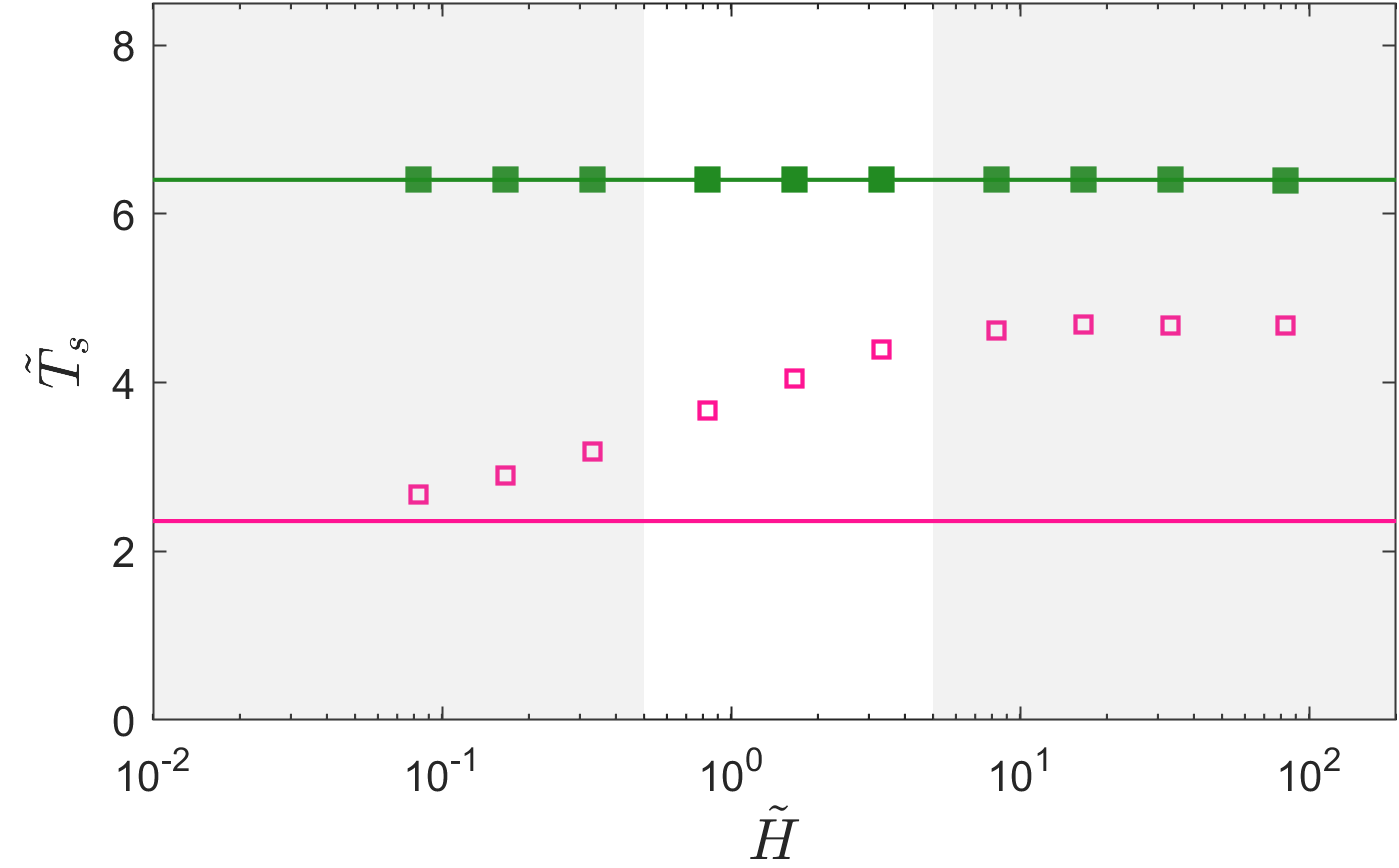}
\caption{ \textbf{ Surface residence times under confinement: memory effects.}
Residence times at the surfaces $\Tilde{T}_s=T_s/\tau_r$ as a function of confinement height $\Tilde{H}$. The white area corresponds to the experimental range of measurements. The P Model outcomes (filled green squares) match the prediction of Eq.~\ref{eq:TS}, in green line. For the BV Model (empty pink squares), the same expression adapted to the BV Model mean run time, in pink line, fails at weak confinement. This is a non-Markovian effect.}
\label{fig4a}
\end{figure}
\section{Memory-induced surface selection}
In the P Model, when a bacterium lands on a surface, the expectation for a tumble is solely defined by the mean run time $\tau_r^P$ and totally independent on the past experiences. This is obviously related to the memoryless nature of a Poisson process. After a tumbling event, the bacterium is bound to leave the surface with a probability $1/2$, and thus the average surface residence time is $\tau_r^P+\tau_t$. Otherwise, the bacterium will dwell on the surface for another average time $\tau_r^P+\tau_t$. At the end of the second tumble, the bacterium will leave the surface with probability $1/2$ and the probability of observing a residence time $ 2(\tau_r^P+\tau_t)$ is $1/4$. Therefore, the probability of observing a residence time $ k(\tau_r^P+ \tau_t)$ is $1/2^k$ which finally yields a mean surface residence time: 
\begin{equation}
\label{eq:TS}
{T}_s^P=\sum_{k=1}^\infty k ({\tau}_r^P+{\tau}_t) \frac{1}{2^k}=2({\tau}_r^P+{\tau}_t)\end{equation}
This theoretical value is displayed in Fig.~\ref{fig4a} for comparison with the numerical outcomes. 
Interestingly, for the BV Model, there is a clear impact of the non-Markovian character of the run time distribution that cannot be captured by the P Model. The presence of an internal memory induces a sensitivity of the surface residence time to the confinement height and a propensity to stay at surfaces longer than what a simple memory-less estimation  would predict [Eq.~\ref{eq:TS}] (note that on Fig.~\ref{fig4a}, $T_s^P$ is presented normalized by $\tau_r$ and not by $\tau_r^p$). Indeed, a bacterium performing long runs (low mood value) will more likely reach a surface and then dwell on it for longer times (see Fig.~S4).
Therefore, the BV Model generically predicts the existence of a sorting process with respect to the distribution of run times. This could have general consequences on the organization properties of bacterial populations. However, currently, it is impossible to test directly this prediction and for the measured $T_s^{*}$ values, the experiments are currently too noisy to identify the predicted (tiny) variations with $H$ [Fig.~\ref{fig3}a].

\begin{figure}[t!]
\centering
\includegraphics[width=1\linewidth]{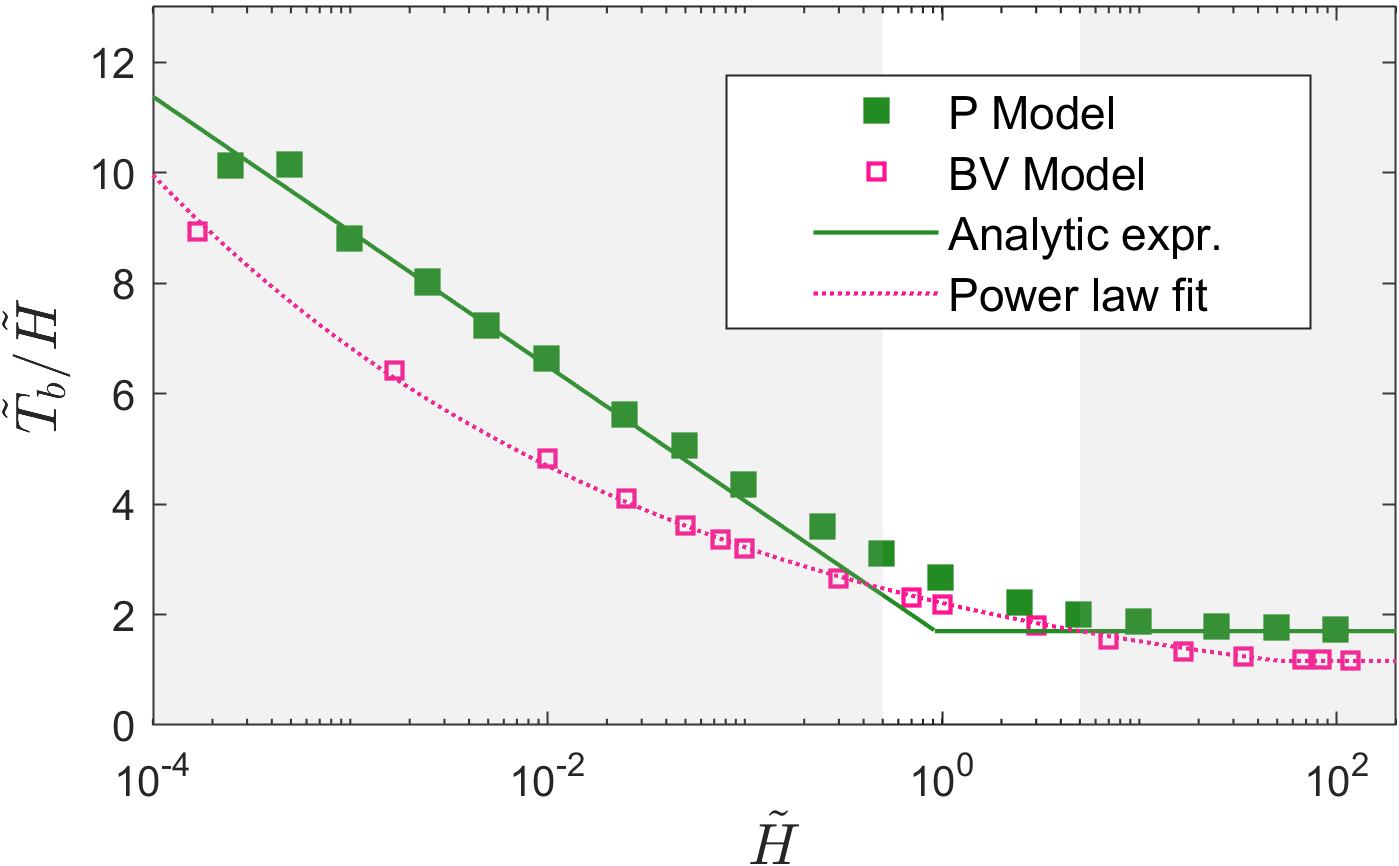}
\caption{ \textbf{ Bulk residence times under confinement: asymptotic behaviors.}
Rescaled residence times in the bulk $\tilde{T}_b/ \tilde{H}$ as a function of the confinement height $\tilde{H}$. The white area corresponds to the experimental range of measurements. P Model predictions (filled green squares) are described in green lines by equation~\ref{eq:SC} (resp.~\ref{eq:LC}) in the regime of strong (resp. weak) confinement. The BV Model outcomes (empty pink squares) are fitted in pink dotted line by a power law of exponent $\beta=-1/6$. Both models show a strong-confinement asymptotic singularity due to the grazing take-off angle contributions.}
\label{fig4b}
\end{figure}
\section{Asymptotic behaviors}
Numerical simulations established that the P Model constitutes a reasonable semi-quantitative description of the emerging lateral dispersion process. The toy model discussed below provides an accurate analytical description of this Markovian model and, importantly, captures the asymptotic scaling of the spreading dynamics at both strong and weak confinement. 
\subsection{Dimension Mixing Diffusive (DMD) model}
The "Dimension Mixing Diffusive " (DMD) model we propose is a simple binary mixture, where "3D-bulk" and "2D-surface" states are treated as independent, with their significance relative to their respective residence times. The diffusion coefficient under confinement is then simply obtained as a weighted average of the unbounded bulk diffusivity $D_{\infty}$ and the surface diffusivity ${D}_{s}(R)$:  
\begin{equation}
\label{eq:D1} {D}({H,R})= \frac{T_b(H) {D}_{\infty} ~+~ T_s^P {D}_{s}(R)}{T_b(H)+T_s^P} \end{equation}
where ${T}_s^P$ is a constant independent of ${H}$ [Eq.~\ref{eq:TS}]. 
By construction, in the limit $T_b \to 0$, the surface diffusivity is recovered, while in the limit $T_b \to \infty$, one obtains the unbounded bulk diffusivity.

\begin{figure*}[t!]
\centering
\includegraphics[width=\linewidth]{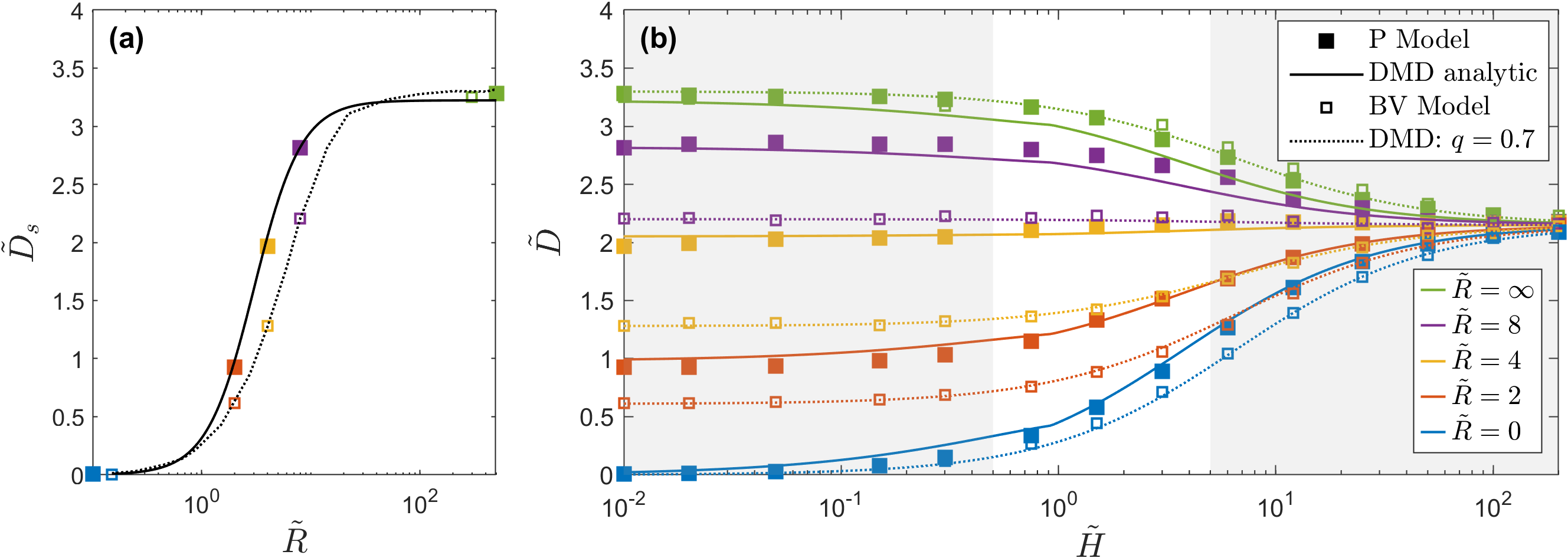}
\caption{ \textbf{Numerical simulations of the P and BV models compared with the analytical expression of the DMD model.}  
P Model (BV Model) predictions are displayed as filled (empty) squares and the corresponding DMD expressions in solid (dotted) lines. Colors correspond to radii of gyration and the white area to the experimental range of measurements.
\textbf{(a)} Surface diffusion coefficient $\tilde{D}_s$ vs. gyration radius $\tilde{R}$. Equation~\ref{D_sP}, displayed in solid line, matches the numerical results of the P Model. Dotted line is the numerical results obtained for the BV model with $\tilde{H}=10^{-4}$.
\textbf{(b)} Spreading diffusivity $\tilde{D}$ vs confinement height $\tilde{H}$, for different radii of gyration $\tilde{R}$ (color-coded). The analytical DMD model [Eq.~\ref{eq:D2},~\ref{eq:D3}] is shown in solid line and compares very well with the P Model, over the whole range of confinement heights $\tilde{H}$ and radii of gyration $\tilde{R}$. Adjusted DMD Model (dotted line) matches all the BV Model outcomes, and thus the experimental data, provided a single adjustment variable $q=0.7$ featuring the ballistic boundary layer extension.}
\label{fig5}
\end{figure*}
\subsection{Weak confinements: diffusive approximation}
According to Eq.~\ref{eq:D1}, the only dependence of the lateral diffusivity $D$ on the confinement height $H$ is through the average bulk residence time ${T}_b$. 
Interpreting ${T}_{b}$ as the time it takes for a bacterium, starting from either the bottom or the top surface, to return to one of these two surfaces, the problem reduces to a mean first passage time (MFPT) problem, computed in the $\hat{z}$-direction. This specific property for a R$\&$T stochastic process has been studied to understand optimal strategies with distributed traps \cite{Rupprecht_2016}. 
For the return to a single surface an analytical expression was even derived in any dimension \cite{Mori_2020}, confirming earlier 1D exact analytical expressions \cite{angelani2015run, Malakar_2018}. Here, we present a simple heuristic argument suited to capture the expected scaling of the bulk residence time. First considering a diffusive 1D process with a bulk diffusion constant $D_{\infty}$, the MFPT to reach either extreme of a segment of size $H$ would be: $T_b = \frac{1}{2 D_{\infty}} z_0 (H - z_0)$, where $z_0$ is the initial position \cite{Redner_2001}.
Obviously, this expression cannot be used with $z_0=0$ (or $z_0=H$) as it would yield the absurd result of a zero MFPT.
However, bacterial motion is not diffusive at all scales. The average run time determines a ballistic regime, and take-off from a surface certainly starts with bacteria in a run phase. 
We then propose to adapt the diffusive MFPT theory and define an effective initial position $z_0=q V \tau_r$, where $q$ is a dimensionless geometric constant of order $1$, interpreted as the contribution of the ballistic boundary layers taking place in the surface vicinity. Replacing this value into the diffusive MFPT approximation (assuming $z_0 \ll H$), yields a relation $ T_b \propto H  $ which is observed numerically at weak confinement for both the P and the BV Models [Fig.~\ref{fig4b}]. The P Model gives explicitly :
\begin{equation}
\label{eq:LC}
H \gg V\tau_r^P, ~ T_b \to \left(\frac{qV^2\tau_r^P}{2D_\infty}\right) \frac{H}{V}
\end{equation}
Note that $D_\infty\propto V^2$, then $T_b \propto H/V$. $T_b$ is not proportional to $H^2$ as could be naively expected for a diffusive process.
Matching asymptotically this expression to the numerical results would give $q \rightarrow q_\infty=1.2(1)$.
However, this approach is bound to fail for confinements reaching $\Tilde{H}=O(1)$.
\subsection{Strong confinements: ballistic approximation}
At stronger confinement, the P Model simulations [Fig.~\ref{fig4b}] display a subdominant logarithmic divergence $ T_b/H \propto \log(H)  $. The BV Model also shows a singularity but probably not logarithmic and better fitted with a power law $T_b/H \propto H ^\beta$ with $\beta = -1/6$ (see Fig.~S5 for a log-log plot). In practice, this anomalous scaling regime is limited by the ultimate confinement i.e. when $H$ reaches a typical bacteria thickness, here $\Tilde{H} =O( 10^{-2})$. Moreover, because of hydrodynamic coupling taking place in this ultra-confined situation (typically $H<5 \mu$m ), the independence of  top and bottom surfaces on the kinematic process cannot be assumed anymore \cite{libberton2019efficiency, tokarova2021patterns, vizsnyiczai2020transition}.

The  logarithmic singularity at high confinement is a generic feature of the grazing take-off angles contribution [Appendix E]. Here, we consider an asymptotic model baptized the "1-run" Model [Appendix F]. This model provides in the limit $H\to 0$ not only an heuristic vision of the transport process but also an analytical expression for the bulk mean residence time without adding free parameters. The central idea is to consider the very first run escaping from a surface and assess its contributions over all possible run times and escaping angles. The key is to separate runs interrupted by the opposite surface from the ones that do not reach it .
This model yields the asymptotic equation :
\begin{equation}
 H\ll V\tau_r^P, ~  T_b \to  \chi \frac{H}{V}
 \label{eq:SC}
\end{equation}
where $\chi=\left[1-\gamma -\ln\left(H/V\tau_r^P\right)\right]/f^P$, $\gamma\approx0.58$ is the Euler constant and $f^P=\tau_r^P/(\tau_r^P+\tau_t)$. This expression compares very well with the numerical results of the P Model [Fig.~\ref{fig4b}]. 
\subsection{Analytical expression}
Bringing together the confined and unconfined limits for the bulk residence times, we propose a general relation encompassing all the regimes. Combining equations~(\ref{eq:TS}),~(\ref{eq:LC}) and~(\ref{eq:SC}), the DMD model [Eq.~\ref{eq:D1}] becomes: 
\begin{equation}
\label{eq:D2} D = D_{\infty} \left[ \frac{q V H f^P +  4D_s}{q V H f^P +  4D_{\infty} }\right]\end{equation}
where $q$ is now a parameter representing subdominant variations of the mean bulk residence time with confinement. Its full evaluation is still theoretically challenging. We then propose, for practical reasons, the explicit approximated analytical expression, that catches both limiting regimes:
\begin{equation}
\label{eq:D3} q = \max \left\{q_{\infty},2\chi {D}_{\infty} / 
 V^2 {\tau}_r^P\right\}
\end{equation}
The DMD model analytical relation is tested against simulations, including the BV Model [Fig.~\ref{fig5}]. The agreement with the P Model is remarkable.

It is worth noticing that the BV Model fails to agree quantitatively with the DMD model when all contributing functions are adapted [Fig.~S5,S6]. 
This is essentially due to the non-Markovian  memory effects that couple non-trivially bulk and surfaces motion across all scales of confinement. Nevertheless, on a qualitative level, the heuristic character of the Mixing Model remains useful in assessing the role of the different contributions.

The DMD model yields under weak confinement: 
\begin{equation}
     H\gg V\tau_r^P, ~ D \to D_{\infty}\left[1+\frac{4}{qVHf^P} \left(D_s-D_\infty \right)\right]
     \label{DweakH2}
\end{equation}
This asymptotic relation highlights a rather slow convergence towards the unbounded limit, hence stressing on the need to explicitly consider geometrical parameters in many problems of microswimmers spreading under confinement.
At stronger confinement --  corresponding mostly to the experimental situation -- it yields:
\begin{equation}
     H\ll V\tau_r^P, ~ D \to D_{s}\left[1+\frac{qVHf^P}{4}\left(\frac{1}{D_s}-\frac{1}{D_\infty}\right)\right]
     \label{DweakH}
\end{equation}
The  dominant scaling is $D \propto H ~V$, which is notably different from the \textit{a priori} expected and often used, scaling relation\cite{Najafi_2018}  $D \propto V^2 \tau_r^P $. This new scaling is quite general as it qualitatively stems from the runs interrupted by surface encounters that induce an effective run time in the bulk $ \propto H/V$, replacing $\tau_r^P$ in the scaling relation.

Though the DMD approach was developed to get theoretical insights on the P Model, we find that this approach -- Eq.~(\ref{eq:D2}) --  can be applied to the BV model with a single adjusted boundary layer value $q = 0.7$ and provided the numerical determination of the corresponding surface  diffusivity $D_S$ [Fig. \ref{fig5}(a)]. Thus, it provides a simple physics-based formula to capture experimental data. The result can be visualized on Fig.~\ref{fig5}(b) and agrees very well with all numerical data. As a consequence, the asymptotic scaling $H \rightarrow \infty$, Eq. (\ref{DweakH2}), derived for the memoryless Poisson process, still holds for the non-Markovian BV model.  

\section{Discussion}
Here, we presented the first experimental study of the long-time spreading dynamics of wild-type Escherichia coli in confinement.  Experimental outcomes  are compared with  a non-Markovian stochastic model (BV-Model) that explicitly accounts for the internal fluctuations of a phosphorylated protein (CheYP) driving the switches of the rotary motors. This model displays a biological memory and induces a large log-normal distribution of the run times. By explicitly accounting for chiral kinematics at surfaces, the dependence of spreading diffusivity and residence times on confinement can be quantitatively reproduced without parametric adjustment. This excellent agreement emphasizes the significance of internal biological fluctuations in the emergence of large-scale transport properties.

Based on these results and using a companion Markovian model (P-Model), we propose a simple dimensional mixing diffusive (DMD) model yielding an approximate expression for the diffusivity that captures the interplay between  confinement and surface kinematics. This conceptual picture is potentially applicable to many other types of active swimmers, including various bacteria strains, unicellular algae, and other motile protists. A combined theoretical and numerical analysis reveals the mechanisms contributing to the confinement-dependent diffusivity in the strong ($H \ll V\tau_r$) or weak ($H\gg V\tau_r$) confinement limits. 
Spreading in strong confinement is controlled by ballistic displacements between the top and bottom surfaces. On the other hand, diffusivity in weaker confinement depends on the average time the micro-swimmer spends in the bulk. We showed that this can be computed as a first-passage time between the bottom and the  upper surfaces, which we quantified by introducing the notion of an effective ballistic boundary layer at the surfaces.
%
We also demonstrate the significant role of take-off dynamics from surfaces. For microorganisms with grazing take-off angles, such as most flagellated bacteria, we expect to see subdominant scaling of the bulk residence time with confinement height. This takes the form of a logarithmic contribution for memory-less swimmers and a weak power law for non-Markovian swimmers.
However, this could differ for other microorganisms, such as swimming algae or paramecia, which interact with surfaces in very different ways.
Finally, our analysis suggests that the convergence towards the bulk boundless limit $D_{\infty}$ scales, generically -- independently of microswimmer details -- as  $D_{\infty} - D(H)  \propto H^{-1}$.  
Therefore, it is not possible to  define a characteristic confinement height beyond which the impact of the walls can be ignored. 
Adjusting the BV-model outcomes quantitatively to the DMD model demonstrates the robustness of the approach.
%
In the case of {\it E. coli} modeled as a non-Markovian swimmer, the confinement height must be nearly $100$ times the characteristic mean run length  to approach significantly the unbounded diffusion limit. This means that the confinement height must be as large as $6000\,\mu\text{m}$ to reach the asymptotic value at less than $5\%$.
Therefore, in most natural environments where more complex or widely distributed confinement conditions are expected, the interplay between surface and bulk motion, which is ultimately modulated by internal biological processes, likely controls bacterial transport properties at (almost)  all scales.
\begin{acknowledgments}
All authors acknowledge the ANR-22-CE30 grant "Push-pull".
F.P. acknowledges additional financial support from C.Y. Initiative of Excellence (grant Investissements d'Avenir ANR-16-IDEX-0008), INEX 2021 Ambition Project CollInt and Labex MME-DII, projects 2021-258 and 2021-297.
\end{acknowledgments}

\appendix

\section{Bacteria preparation}
Strain AD62 is pre-cultured overnight (about 14 hours) in Lysogeny Broth Lennox (LB). Bacteria are then cultured for 4 hours by diluting the pre-culture 1:50 in Tryptone Broth (TB : 10g/L tryptone + 5g/L NaCl). OD reaches 0.5.
The resulting solution is then centrifuged (5 minutes at 4590 RCF) and the supernatant medium is replaced by Berg's Motility Buffer (BMB : 3.9 g/L NaCl + 0.1mM EDTA + 25g/L L-Serine + 10mM phosphate buffer 100mM), a minimal medium preventing from bacterial growth. Bacterial solution is then diluted (OD = $1.10^{-4}$) to avoid collision during tracking experiments and mixed 1:1 with Percoll (P1644) so bacteria and the surrounding fluid have the same density (d = 1.06).
\section{Raw experimental kinematic parameters}
For each 3D tracking experiment labeled $ {\kappa}$, kinematic parameters such as swimming velocity $V^{\kappa}$ , radius of gyration $R^{\kappa}$ and diffusivity $D^{\kappa}$ are  extracted. Also, the tracks provides a precise measurement of the actual confinement height $H^{\kappa}$.
The procedure for the extraction of the parameters is detailed below and shown in  Fig.~S2. Statistics on the population are discussed in  Fig.~S3.
\subsection{Swimming velocity $V$}
The instantaneous velocity $\textbf{v}$ is obtained by smoothing the trajectory using a Savitsky-Golay algorithm of order 2, in which we define the velocity as the derivative of the Savitsky-Golay polynomial output. The smoothing time for the in-plane coordinates $x$ and $y$ is chosen as $0.2s$ whereas the smoothing time for the out-of-plane z coordinate is chosen as $0.5s$.
The distribution of the instantaneous speed $v$ shows two Gaussian contributions. One is coming from tumble events (lower speed) and the other from the runs (higher speed). 
We define the swimming speed $V$ as the mode of the run contribution, fitted by a Gaussian distribution. The fit is done on the values of the speed above half-maximum.
\subsection{Surface radius of gyration $R$}
In near-surface swimming, the radius of gyration of an individual bacterium is mostly constant, but can vary a lot within the bacterial population.
A trajectory can be considered as a sequence of successive bulk and surface "sub-pieces". To access the surface radius of gyration, a virtual trajectory is built by aggregating all the surface "sub-pieces". To avoid taking into account trajectories close from surfaces but actually in the bulk, we consider as "surface pieces" the parts closer than $3$ microns from a surface. From this artificial trajectory is computed its associated orientation auto-correlation function :
\begin{align}
    \left< \cos \theta (\Delta t) \right>&=\left<  \frac{\textbf{v}(t+\Delta t) . \textbf{v}(t)}{\lVert \textbf{v}(t+\Delta t) . \textbf{v}(t) \rVert} \right>\\
    &=\frac{1}{T-\Delta t}\int_0^{T-\Delta t} \frac{\textbf{v}(t+\Delta t) . \textbf{v}(t)}{\lVert \textbf{v}(t+\Delta t) . \textbf{v}(t) \rVert} \,dt
    \label{eq:autocorr}
\end{align}
The resulting quantity displays damped oscillations, as a result of orientational noise and circular motion, that can be fitted by a function $f_{ACF}$ decaying exponentially on a characteristic persistent time $\tau_p$, in a sinusoidal envelope of pulsation $\Omega$ :
\begin{equation}
    f_{ACF} (\Delta t)=e^{- \Delta t/\tau_p}\cos \left( \Omega \Delta t \right)
    \label{eq:ACF}
\end{equation}
Note that the determination of the pulsation is more robust at short time lags $ \Delta t < \tau_r$, where the impact of tumbles on the reorientation is statistically less significant than the signature of the circular motion. The fit is then done on the two first seconds only.
Finally, $R=V/\Omega$.
\subsection{Lateral diffusivity $D$}
The lateral diffusion coefficient characterizing the in-plane spreading is extracted from the self-averaged mean square displacement $\Delta \textbf{r}^2 (\Delta t)$ computed on the trajectory projected on the XY plane of interest $\textbf{r(t)}=(x(t),y(t))$ : 
\begin{align}
     \left< \Delta \textbf{r}^2 (\Delta t) \right>&=\left< \left[ \textbf{r}(t+\Delta t) - \textbf{r}(t) \right]^2 \right>\\
     &=\frac{1}{T-\Delta t}\int_0^{T-\Delta t} \left[ \textbf{r}(t+\Delta t) - \textbf{r}(t) \right]^2 \,dt
    \label{eq:MSD}
 \end{align}
The mean square displacement is fitted with F{\"u}rth's formula, which depends on two parameters, the ballistic speed and the characteristic ballistic time $\tau_c$: 
 \begin{align}
    f_F(\Delta t) &=2 V_c^2 \tau_c^2 \left( \frac{\Delta t}{\tau_c}- 1 + e^{-\Delta t/\tau_c} \right)\\
    \label{eq:Furth}
    &\underset{\Delta t \to 0}= V_c^2 \Delta t^2\\
    &\underset{\Delta t \to \infty}= 2 V_c^2 \tau_c \Delta t = 4D\Delta t
\end{align}
The MSD is fitted on a time lag $\Delta t=T/10>20s$, as a compromise between the statistical convergence of self-averaged quantities and the importance to probe the diffusive regime, i.e lag times $\Delta t > \tau_c \approx \tau_r=2.23s$. We remind that $T$, the duration of the experimental track, lies between 200s and 300s, then indeed $T/10>\tau_c$.
Finally, $D = V_c^2 \tau_c / 2$
\subsection{Height of the pool $H$}
The height of the pool (i.e. the distance between the top and bottom surfaces) is roughly chosen with the choice of spacers made of double face tape. The actual separation can vary from one experiment to the other depending on the actual compression of the tape. A precise estimate is obtained through the distribution of the position along the vertical z-axis of a trajectory. Indeed, since the bacteria spend long times at surface, the location of bottom and top surfaces are revealed by two peaks in this distribution, that can be fitted with a double Gaussian. Thus, bottom and top locations are measured precisely for each experiment.
\section{Residence times $T_b^*$ and $T_s^*$}
Residence times are obtained using the same double-threshold method as Junot et al. \cite{Junot2022}: Below the surface threshold $\delta_s=3\mu m$ the bacterium is considered at surface. Above the bulk threshold $\delta_b=8\mu m$ it is considered in the bulk. In between, the status is determined a posteriori: it is considered at surface only if it comes from the surface and reaches the surface threshold $\delta_s$ again without crossing the bulk threshold $\delta_b$ (else, it is considered in the bulk).
\section{Numerical models}
\subsection{Behavioral Variability (BV) Model}
The swimming velocity is $\mathbf{V} = V \hat{\mathbf{p}}$, where $\hat{\mathbf{p}}$ is a unit vector that denotes the swimming direction and $V$ is the swimming speed, considered as constant both, in the bulk and at surfaces.
During a run phase, the swimming direction remains constant in the bulk, whereas at surfaces the moving direction first is set parallel to the surface and then evolves as  
$\dot{\hat{\mathbf{p}}}= \epsilon \Omega \hat{\mathbf{z}} 
 \times  \hat{\mathbf{p}}$
 where $\Omega$ 
is the angular rotation speed  and $\epsilon$ is set to $\epsilon=1$ (resp. $-1$) at the bottom (resp. top) surface. Thus, the swimmer undergoes circular trajectories of radius $R=V/\Omega$.
The  run-phase duration -- independently whether the bacterium is in the bulk or at a surface -- is controlled by the rate $\nu = \nu_0 \exp[\Delta_0 X(t)]$, which depends on the internal variable $X(t)$ (related to the concentration of CheY-P proteins). The stochastic variable $X(t)$ obeys an Orstein-Uhlenbeck process : $\dot{X} = - X/\tau_{M} + \sqrt{2/\tau_{M}}\xi(t)$, where $\xi(t)$ is a Gaussian noise and $\tau_{M}$  a memory time.
During the tumbling phase, the swimming velocity is  $V=0$ and $\hat{\mathbf{p}}$ is randomly reoriented through a diffusive process characterized by a rotational diffusion constant $D_r$. 
The duration of a tumble event is taken from a Poisson distribution of mean time $\tau_t$.
The tumbling process remains the same, both in the bulk and at the surfaces. 
The bulk-to-surface transition  is simply determined by an alignment rule \cite{Junot2022}. When the bacterium reaches either the top or the bottom surface, its orientation aligns instantaneously to the surface and a circular rotation starts tangentially to this initial orientation. The transition from surface to bulk occurs if, after tumbling, the bacterium orientation points towards the bulk, otherwise the bacterium is realigned parallel to the surface and a circular motion resumes. 
\subsection{BV Model simulation parameters}
BV Model parameters are : $V = 27 \mu m/s$, $\tau_{M}=19 s$ , $\Delta_0=1.62$ and $\nu_0=0.216\,s^{-1}$. Extensive simulation of the stochastic process  yields a mean run time  $ \tau_r = 2.23\, s$ and a second moment parameter $ S = 3.5$.
Tumbling times are drawn from a Poisson distribution with a mean 
tumbling time  $\tau_t = 0.4\, s$. The reorientation process during tumble is a 3D orientation diffusion with a dimensionless diffusivity $\Tilde{D}_r = D_r \tau_t = 3.86 $.

\subsection{Boundless asymptotic limits $D_{\infty}$}
Simulations of the run and tumble BV Model in absence of boundaries yield, at large time, a diffusive process characterized by a diffusion coefficient $\Tilde{D}_{\infty}=D_{\infty}/D_0=2.1485$.
To compute this values using Eq.~\ref{LD}, we need to estimate the mean reorientation during a tumble event, i.e. the value of $\alpha$. 
For a rotational diffusive process in \textit{d}-dimension, characterized by an angular diffusion coefficient $D_r$, the mean orientation de-correlation after a time $t$ is $\exp{[-(d-1)  {D}_r t]}$. 
Hence, for a Poisson distribution of tumble times $P(t) = 1/\tau_t \exp{[-t/\tau_t]}$, the average mean cosine is: $\langle {\cos{\theta}} \rangle = \int_{0}^{\infty} \frac{1}{\tau_t} \exp{[-t/\tau_t]} \exp{[-(d-1)  {D}_r t]} dt$. We obtain: 
\begin{equation}
1-\alpha^{3D}= \frac{2\Tilde{D}_r}{1+2\Tilde{D}_r}  
\end{equation} 
This result is consistent with the numerical simulations and corresponds to  $\alpha^{3D}= 0.1147$. 
Markovian character of the BV-model.\\

\subsection{P Model simulation parameters}
To obtain an asymptotic match between BV Model and  P Model, provided a Poisson run time $\tau^{P}_{r}$, one should equate the diffusion coefficients for $H\to\infty$:  $\frac{V^2 \tau^{P}_{r}}{3(1-\alpha_{3D})}\frac{\tau^{P}_{r}}{\tau^{P}_{r}+\tau_t}=D_{\infty}$, where $D_{\infty}$ is obtained using equation \ref{LD} plugging in the parameter obtained to simulate the BV Model. 

The kinematics of the P Model for the run and tumble phases is similar to that of the BV Model. The only difference lies in the mean run time $\tau_r^P=6.74\,s$, which remains constant.
\subsection{Confined asymptotic limits $D_s$}
For a P Model taking place only at the surface ($\Tilde{H} \rightarrow 0$) with pieces of circular trajectories of radii $R$, the  corresponding diffusion coefficient can be derived exactly.
The asymptotic diffusion coefficient can be obtained by adding jumps of value $\Vec{l_i}$ with a magnitude $\lVert \Vec{l_i} \rVert = 2R |\sin{(\theta_i/2)}|$, with $\theta_i$ the angular rotation of the swimmer on the circle from the moment the circular motion begins to the end of the rotation when tumble starts. Using the Lovely and Dahlquist expression, the diffusion constant at the surface is then 
$D_S =\frac{<l^2>}{4 \tau^P_r} \frac{f_P} {(1-\alpha(R))}$
$=\frac{R^2}{\tau^P_r}  \int_{0}^{\infty} \sin^2{(\frac{V_B \tau}{2R})} \exp{[\frac{-\tau}{\tau^P_r}]}\, \frac{f_P} {(1-\alpha(R))} d\tau $ with $f_P = \frac{\tau^P_r}{(\tau^P_r+ \tau_t)}$.  Finally, it is  obtained: 
\begin{equation}
\Tilde{D}_s =  \frac{\Tilde{D}_s^{\infty}}{1+(\Tilde{R}~\tau_r/\tau_r^P)^{-2} } 
\end{equation} 
with 
$\Tilde{D}_s^{\infty}=\frac{\tau^P_r}{\tau_r} \frac{ f_P} {(1-\alpha(R))}$.
There is a disucssion on $\alpha(R)$. When $R\ll V \tau_r^P$ then there is random reorientation thus $\alpha(0)=0$ whereas it is maximal when $R\to \infty$ and $\alpha(\infty)=\alpha^{2D}$. Note that $\alpha^{2D}$ is not stemming from a strict 2D reorientation process but comes from a 3D diffusive reorientation and then, a projection on the surface. It can be found numerically and its value is $\alpha^{2D}=0.13(1)$, leading to $\Tilde{D}_s^{\infty}=3.28(2)$. Using the approximate value $\alpha^{2D}=\alpha^{3D}=0.1147$, leads to $\Tilde{D}_s^{\infty}=3.22(2)$, very close from the exact value, as discussed in Saragosti et al \cite{Saragosti2012}. Then we assume for simplicity: $\alpha=\alpha^{3D}$ for all $R$ and thus $\Tilde{D}_s^{\infty}=3.22(2)$.
\section{Logarithmic dependence in $T_b$ ($\tilde{H}\ll 1$)}
For any value of $H$, independently of how small is the gap, we can always find a take-off angle $\theta$ -- between the surface and the bacterium velocity -- such that the speed in $\hat{z}$ is $V_z = V \sin(\theta)$ results in $T_b(\theta)=H/V_z \gg \tau_r$. 
Moreover, $T_b$ is not a simple linear function of $H$, there exists logarithmic corrections. 
To understand this, let us first note that $T_b = \int_0^{\pi} T_b(\theta) d\theta$ diverges. 
To avoid this divergence, we include a cut-off angle $\theta^*$ computed considering that the average run length of the bacterium is $V \tau_r$, and thus, $\sin(\theta^*)=H/(V \tau_r)$. 
Furthermore, we assume that grazing angles -- those in the intervals $[0, \theta^*]$ and $[\pi-\theta^*, \pi]$ -- contribute with bulk times of the order of $\tau_r$: after taking off with a grazing angle and moving a distance $V \tau_r$, the reorientation resulting from the tumbling event will allow the bacterium to quickly reach on of the surfaces. 
Putting all together, and for simplicity assuming that all take-off angles exhibit the same probability, i.e. $p(\theta)=1/\pi$, we obtain: 
\begin{align}
\label{eq:Tb} 
T_b(H) \!&=\!\frac{1}{\pi} \int_{\theta^*}^{\pi-\theta^{*}}\!\!\! T_b(\theta) d\theta \!+\! \frac{2 \theta^* \tau_r}{\pi}  \! \nonumber\\
&=\! \frac{2 H}{\pi V}\! \left[ 1\! - \ln\! \left(\frac{H}{2 \tau_r V} \right) \right]  
\end{align}
Despite the many approximation performed to arrive at expression~(\ref{eq:Tb}), 
it displays the correct functional form, in particular it explains that 
$T_b(H)/H \simeq - m\,\ln(H) + b$ with $b$ and $m$ constants. 
\section{"1-run" Model  ($\tilde{H} \ll 1$)}
This model aims at capturing the lateral diffusion coefficient $D$ and the residence time in the bulk $T_b$ in the strong confinement regime (i.e $H\ll1$) by considering the bacterium goes from one surface to the other one in one run phase. 
These two quantities are averaged over all  take-off angles with probability $P_\theta(\theta)$ and all run times with probability $P_T(t)$. For the P Model $P_\theta(\theta)= \cos(\theta)$ (equal probability on the sphere) and $P_T(t)=\exp(-t/\tau_r^P)/\tau_r^P$, but the method is quite general: other distributions for different models could be considered. Two different contributions are taken into account: 
i) the runs going directly from one surface to the other one, and
ii) the runs that end in the bulk without reaching the opposite surface. 
These two types of run are treated independently, yielding two different terms in the integral.
Here are the expressions and results for strong confinement. 
For the residence time in the bulk, it comes :
\begin{align}
{T}_{b}({H})
&=\int_0^{\pi/2}\left\{\int_0^{t_{max}(\theta,H)} \left[t+\tau_t\right] P_T(t) dt \right.\nonumber\\ & 
+ \left.\int_{t_{max}(\theta,H)}^{\infty} \left[t_{max}(\theta,H)\right] P_T(t) dt \right\} P_\theta(\theta)d\theta\nonumber\\
&\underset{0}{=} (1+\tau_t/\tau_r^P)(1-\gamma -\ln({H}/{V\tau}_r^P))H/V
\end{align}
where $t_{max}(\theta,H)=H/Vsin(\theta)$ and $\gamma\approx0.58$ is the Euler constant, which is Eq.~\ref{eq:SC}.
For the diffusion coefficient, a solution can be derived, based on the computation of run lengths \cite{Lovely1975}, though it is not clear that it can be applied under confinement.
In the case of small radii of gyration $\tilde{R}\ll1$ (in this case no diffusion occurs at surface and the orientation of the particle is randomized at surface), it yields :
\begin{align}
    D(H)&=\frac{1}{2d}\frac{\left<\ell^2_{2D}(H)\right>}{T_{b}(H)}\frac{T_{b}(H)}{T_{b}(H)+T_s^P} \\
    &\underset{0}{=} \frac{1}{2d}\frac{\left<\ell^2_{2D}(H)\right>}{T_s^P}
\end{align}
where $\ell_{2D}=\ell cos(\theta)$ is the length of the projection of a run of length $\ell=Vt$ and forming an angle $\theta$ with the XY-plane. We can compute $\left<\ell^2_{2D}(H)\right>$ as :
\begin{align}
    \ell^2_{2D}(H)
    &=\int_0^{\pi/2}\left\{\int_0^{t_{max}(\theta,H)} \left[\ell(t)\cos(\theta)\right]^2 P_T(t) dt \right.\nonumber \\
    &+ \left. \int_{t_{max}(\theta,H)}^{\infty} \left[\ell_{max}(\theta,H)\cos(\theta)\right]^2 P_T(t) dt \right\} P_\theta(\theta)d\theta \nonumber\\
    &\underset{0}{=} 2 VH\tau_r^P
\end{align}
where $\ell_{max}(\theta,H)=Vt_{max}(\theta,H)$.
Finally, it provides the following expression for the lateral diffusion coefficient at strong confinement and for small radii of gyration :
\begin{equation}
    D(H)\underset{0}{=} VHf^P/4
\end{equation}
Remarkably, this expression is excellent to describe quantitatively the diffusivity, for $H \to 0$ and $D_s \to0$, observed in the P Model [Fig.~S7]. Moreover, we recover Eq.~\ref{DweakH} with simply $q=1$, close from the opposite limit value $q_{\infty}$.
Then a simplification of the DMD model consists in taking $q=1$ in Eq.~\ref{eq:D2} for all confinements, in quantitative agreement with the P Model [Fig.~S8].

\bibliography{Bibliography}

\end{document}